\documentclass[aps,prapplied,preprint,superscriptaddress]{revtex4}

\usepackage{amssymb}
\usepackage{yfonts}
\usepackage{graphicx}
\usepackage{amsmath}
\usepackage{color}
\usepackage{epstopdf}
\usepackage{ulem}

\begin{document}

\title{Controllable dispersion of domain wall movement in antiferromagnetic
thin films at finite temperatures}
\author{Yuriy G. Semenov}
\affiliation{Department of Electrical and Computer Engineering, North Carolina State
University, Raleigh, NC 27695, USA}
\author{Xinyi Xu}
\affiliation{Department of Electrical and Computer Engineering, North Carolina State
University, Raleigh, NC 27695, USA}
\author{Ki Wook Kim} \email{kwk@ncsu.edu}
\affiliation{Department of Electrical and Computer Engineering, North Carolina State
University, Raleigh, NC 27695, USA}
\affiliation{Department of Physics, North Carolina State University, Raleigh, NC 27695,
USA}

\begin{abstract}

The dynamics of a 90$^{\circ }$ domain wall in an antiferromagnetic nanostrip
driven by the current-induced spin-orbital torque
are theoretically examined in the presence of random thermal fluctuations.
A soliton-type equation of motion is developed on the basis of energy balance
between the driving forces and dissipative processes in terms of the domain
wall velocity.  Comparison with micromagnetic simulations in the deterministic
conditions shows good agreement in both the transient and steady-state transport.
When the effects of random thermal fluctuations are included via a stochastic treatment,
the results clearly indicate that the dispersion in the domain wall position can be
controlled electrically by tailoring the strength and duration of the driving current
mediating the spin orbital torque in the antiferromagnet.  More specifically,
the standard deviation of the probability distribution function for the domain wall
movement can be tuned widely while maintaining the average position unaffected.
Potential applications of this unusual functionality include the probabilistic
computing such as Bayesian learning.
\end{abstract}

\maketitle







\section{Introduction}

Effective control of the domain walls (DWs) in the antiferromagnetic (AFM) materials remains a vital
challenge for the prospective spintronic applications \cite{Gomonay2017,Baltz2018}.
A number of approaches explored thus far have demonstrated the promise. For instance, the effect
of the spin-transfer torque was able to manipulate the AFM DWs \cite{Hals2011,Swaving2011,Tveten2013}.
Similarly, the AFM DW transfer can be
induced as a result of the spin-wave mediated forces \cite{Kim2014,Tveten2014},
which are enhanced by the Dzyaloshinskii-Moriya interaction \cite{Qaiumzadeh2018}.
In fact, the Dzyaloshinskii-Moriya interaction also assists the DW
movement in the presence of a rotating magnetic field \cite{Pan2018}.
Further, the normally ineffective magnetic field was shown
to control the AFM DW velocity when its spatial gradient is combined with a charge current
or more precisely, the spin transfer effect \cite{Rodrigues2017,Yamane2017}.
However, the most accessible mechanism for the control of DW dynamics
in the AFM thin films appears to be the spin-orbital torque (SOT).
The electric current through a non-centrosymmetric antiferromagnet generates a SOT associated
with a staggered (or N\'{e}el-order) field \cite{Gomonay2016,Yu2018}.
The DW motion can also be modulated by the SOT induced at the interface with a strongly
spin-orbit coupled material such as a heavy metal \cite{Shiino2016}.

Specific physical details aside, all of the aforementioned studies have considered
the deterministic dynamics essentially in the limit of zero temperature.
Furthermore, most of them have focused on the AFM textures in the form
of 180$^{\circ }$ DWs, posing an additional challenge in discriminating the
domains with antiparallel N\'{e}el vectors. A more promising alternative
may be to utilize the 90$^\circ$ DWs in light of the recent experimental advances
in their manipulation and detection \cite{Yu2018}. In fact, 90$^\circ$ rotation
of the N\'{e}el-vector orientation has already been identified in the electrical measurements
with a large signal in the anisotropic magnetoresistance \cite{Yu2018} or
its tunneling variety \cite{Park2011}.  On the other hand, the theoretical
understanding on the dynamics of 90$^{\circ }$ DWs in the antiferromagnets has received only limited
attention~\cite{Gomonay2016}.  Since the operation is predominantly at the ambient conditions,
the effect of finite temperature and subsequent stochastic nature
of the DW dynamics also add interesting perspectives.


The uncertainty in the final DW position is evidently undesirable for
memory or Boolean logic applications. On the other hand, controllable
stochasticity of the output signals with a desired distribution function has begun to
draw much attention since its prospective utilization in machine learning
or Bayesian computing. Physical implementation of the concept of probabilistic
computing has so far relied predominantly on the external generators of  random
numbers.  While the alternative approaches have also been explored, they tend to involve
complex hardware arrangements (see, for instance, Ref.~\cite{Pinna2018} and the
references therein).  It is clearly desirable if a single device or structure can
meet the needs with simple electrical control.  The probabilistic effect
of the AFM DW motion may offer this unique functionality.

In the present work, we theoretically analyze the 90$^{\circ }$ DW motion in an AFM thin film
driven by the SOT pulse at finite temperatures.  A soliton-type treatment is developed
from the Lagrangian representation to examine the DW dynamics including the
effect of random thermal fluctuations.  The results clearly illustrate the characteristics
of the AFM DW motion.  When subjected to a SOT pulse, the DW undergoes acceleration and
then velocity saturation as the dissipation compensates the anti-damping torque.
Once the driving force is turn off, the inertial phase follows much like an actual
particle with non-zero mass.  The random thermal fields introduce substantial
deviations in the DW trajectories depending on the strength and the duration of
the driving SOT pulse.  The corresponding variation in the probability distribution of
the DW position indicates a broad range that can be tuned electrically.

\section{Theoretical formulation}
The structure under consideration is shown in Fig.~1(a).  It is essentially
a bilayer structure of an AFM nanostrip and a strongly spin-orbit coupled
material that can benefit from the current induced SOT mentioned earlier \cite{Shiino2016}.
A vertical tunnel junction can be added on top for electrical measurement
of the DW position via the tunneling anisotropic magnetoresistance.
As the $x$-directional current flow induces the
effective field along the $z$ axis in the present set-up, the magnetic
domains need to be oriented normal to this axis ($z$) for efficiency. Accordingly,
an antiferromagnet with biaxial anisotropy in the $x$-$y$ easy plane
(such as those with tetragonal $D_{4h}$ symmetry) is assumed for the
desired 90$^\circ$ DW. The metallic materials are expected to offer a
convenient choice for experimental realization and detection, while the similar
physical principles can be applied to the dielectric systems as well.

In the analysis, the relatively small cross-section of the AFM strip makes it possible to ignore
the N\'{e}el-vector variation in the $y$-$z$ plane, reducing the problem
to the dynamics in the 1+1 space-time ($x$,$t$) coordinates with no additional
variables. Since the N\'{e}el-vector reorientation is only in the $x$-$y$ plane, the calculation
can take advantage of the Lagrangian representation in terms of the azimuthal
angle $\varphi (x,t)$ between the two easy axes (using the $x$ as the reference).
A similar formalism developed earlier for an easy-plane antiferromagnet  \cite{Semenov2017}
can be directly generalized to the case of biaxial anisotropy with the following
term for the anisotropy energy,
\begin{equation}
W_{in}=\frac{K}{4}(n_{x}^{4}+n_{y}^{4})=\frac{K}{8}\sin
^{2}2\varphi .  \label{Win}
\end{equation}%
Here, $K$ ($< 0$) is the anisotropy constant and $
n_{x(y)}=L_{x(y)}/\left\vert \mathbf{L}\right\vert = \cos \varphi$ ($ \sin \varphi $) is
the normalized representation $\mathbf{n}$ of the N\'{e}el vector $\mathbf{L}$.  The
$z$-component ($n_z$) does not appear in the dynamical equations due to the hard nature
of this direction \cite{Semenov2017}. Then, the
dimensionless Lagrangian $\textswab{L}=\textswab{L}/\left\vert K\right\vert$ can be
reduced to the canonical form in terms of dimensionless time $t$ ($=\omega
_{r}t$) and space $x$ ($=\omega _{r}x/v_{m}$); i.e.,
\begin{equation}
\textswab{L}=\frac{1}{2}\dot{\varphi}_{(t)}^{2}-\frac{1}{2}\dot{\varphi}%
_{(x)}^{2}-\frac{1}{8}\sin ^{2}2\varphi ,  \label{La}
\end{equation}%
where $\dot{\varphi}_{(i)}\equiv \frac{\partial }{\partial i}\varphi $ ($i = x, t$),
$\omega _{r}=\gamma \sqrt{2H_\mathrm{ex}H_\mathrm{an}}$ denotes the zero-field AFM resonance
frequency, $\gamma$ the gyromagnetic ratio, $H_\mathrm{an}=\left\vert K\right\vert/\frac{1}{2}L$
the anisotropy field, $H_\mathrm{ex}$ the exchange field between the magnetic sublattices,
and $v_m$ the magnon velocity.

The corresponding Euler-Lagrange equation for an open system subjected to
the external forces (i.e., torques) $\rho =\rho (x,t)$ reads%
\begin{equation}
\ddot{{\varphi }}_{(tt)}-\ddot{ {\varphi }}_{(xx)}+\frac{1}{4}\sin 4\varphi =\rho .  \label{EL}
\end{equation}%
It is remarkable that Eq.~(\ref{EL}) reproduces the sine-Gordon equation
in terms of variable $4\varphi $ for a conserving system (i.e., when $\rho =0$). In this case,
the basic solution of Eq.~(\ref{EL}) can be expressed in
the form of a soliton moving with velocity $v$,
\begin{equation}
\varphi =\pm \arctan \left[ \exp \left( \frac{x-vt}{\sqrt{1-v^{2}}}\right) %
\right]  . \label{be}
\end{equation}%
The effective width $\delta_\mathrm{dw}$ of the soliton (or the DW) can be estimated as
$ \pi \sqrt{1-v^{2}}$. The corresponding N\'{e}el-vector texture $\mathbf{n}=(\cos \varphi ,\sin \varphi ,0)$
is shown in Fig.~1(b).  Note that the velocity $v$ is also normalized to $v_m$ following
the definitions of $x$ and $t$ (thus, $\left\vert v\right\vert <1$).

Similarly to Ref.~\cite{Shiino2016}, the soliton representation [Eq.~(\ref{be})] of
the DW texture is treated as an ansatz for the analysis of a non-conserving system with energy
dissipation and external forces. In such an approximation, only the DW velocity $%
v=v(x,t)$ remains the actual parameter describing the DW dynamics. To proceed further, the specific
form for $\rho$ is needed in Eq.~(\ref{EL}).  The relevant contributions in the present analysis
come from three terms$-$the energy dissipation via damping, the anti-damping SOT from
the driving current, and the field-like torque from the random thermal fluctuations.
The resulting expression for $\rho (x,t)$ is given in dimensionless units as~\cite{Semenov2017}
\begin{equation}
\rho =-2\lambda \dot{\varphi}_{(t)}+\Phi (x,t)-\frac{\partial h_\mathrm{th}}{%
\partial t} \,,  \label{rhp}
\end{equation}%
where $\lambda $ ($ = \delta_r / \omega_r$) is the damping parameter related to the
width $\delta_r$ of the AFM resonance and $h_\mathrm{th}$ ($=\gamma H_\mathrm{th}/\omega _{r}$) is the
normalized thermal field $H_\mathrm{th}$. In addition, the second term representing
the anti-damping SOT of $\eta (\mathbf{n\times } \hat{\mathbf{z}})\cdot \dot{%
\mathbf{n}}_{(t)}$ takes the form
\begin{equation}
\Phi (x,t)=\frac{\eta (x,t)}{M_{L}^{2}}\frac{H_\mathrm{ex}}{2H_\mathrm{an}} ,  \label{SOT}
\end{equation}%
where $M_L$ denotes the sublattice magnetization ($\simeq \frac{1}{2} L$) and the parameter $\eta $
reflects the strength of the spin-Hall current flowing into the AFM layer
\cite{Shiino2016}.  $H_\mathrm{th}$ is a natural source of randomness in the DW trajectories
and the dispersion in its final location.  With corresponding modifications to Eq.~(\ref{rhp}),
the formalism discussed above can also be applied to the DW transfer driven by the N\'{e}el SOT in
the asymmetric antiferromagnets \cite{Gomonay2016}.

The dynamics of the N\'{e}el-vector textures are first examined in the absence of thermal
fluctuations, where the current induced SOT unambiguously determines the DW motion.
Applying the ansatz of Eq.~(\ref{be}) sufficiently simplifies the problem to the balance
of energy flows to and from the DW, avoiding the need to solve Eqs.~(\ref{EL}) and (\ref{rhp}) directly.
In this context, the net mechanical energy of the DW with the Lagrangian given in Eq.~(\ref{La}) becomes
\begin{equation}
E(v)=\int \left( \frac{1}{2}\dot{\varphi}_{(t)}^{2}+\frac{1}{2}\dot{\varphi}%
_{(x)}^{2}+\frac{1}{8}\sin ^{2}2\varphi \right) dx.  \label{Ee}
\end{equation}%
The range of integration can be safely set for the entire $x$-axis provided that
the structure is much longer than the DW width. Then, for a
single 90$^{\circ }$ wall of Eq.~(\ref{be}), the calculation yields
\begin{equation}
E(v)=\frac{1}{2\sqrt{1-v^{2}}} ,  \label{Ev}
\end{equation}%
which corresponds to the $1/16$ of the energy for the 360$^{\circ }$ kink described by
the sine-Gordon equation. The description in conventional energy units can be restored by
multiplying Eq.~(\ref{Ev}) by the factor $AKv_{m}/\omega _{r}$ and replacing
$v$ with $v/v_m$.  Here, $ A $ denotes the cross section of the AFM thin film.
Along with Eq.~(\ref{Ev}), the rate of net energy change (${dE}/{dt}$) can also be
obtained from Eq.~(\ref{rhp})
as $\int [\Phi - 2\lambda \dot{\varphi}_{(t)}] \dot{\varphi}_{(t)}dx$. After
some algebra, the balance of energy flow to and from the AFM layer can be
written finally in terms of an equation for the DW velocity%
\begin{equation}
\frac{1}{2\lambda }\frac{dv}{dt}=\left( -v+Q\sqrt{1-v^{2}}\right) (1-v^{2}),
\label{ve}
\end{equation}%
where $Q=\frac{\pi }{2\lambda }\Phi $. The solution of this expression provides
the trajectory of the DW in the deterministic transport (i.e., no thermal fluctuations).

The equation of motion for the DW can be solved analytically under simple conditions.
Assuming the initial stationary state $x=0$ at $t=0$ and a constant SOT $Q$ applied for a
duration $t_p$, Eq.~(\ref{ve}) yields
\begin{equation}
v_{Q}(t )=\frac{Q(1-e^{-2\lambda t })}{\sqrt{Q^{2}(1-e^{-2\lambda t })^{2}+1}},~~0<t
\leq t_{p}.  \label{vt}
\end{equation}
The DW velocity increases linearly as $v_{Q}(t)= 2 \lambda Q t =\pi \Phi t$ in the
beginning stage and then shows a saturation pattern once the dissipation compensates
the anti-damping SOT as $t \gg 1/2 \lambda$,
\begin{equation}
v_{s}=\frac{Q}{\sqrt{Q^{2}+1}}.  \label{vs}
\end{equation}
Note that this steady-state velocity $v_{s}$ reaches the maximal magnon velocity $v_{m}$ (=1)
in the limit of $\lambda \rightarrow 0$ or $Q\rightarrow \infty $.  Another interesting
observation from Eq.~(\ref{ve}) is that the velocity does not drop to zero even
after the driving force is turned off.  With the initial velocity of $v_p$ at $t = t_p$
[i.e., $v_p=v_{Q}(t_p)$], the solution illustrates the characteristic inertial motion with
\begin{equation}
v_{0}(t)=\frac{v_{p}}{\sqrt{e^{4 \lambda(t -t_{p})}(1-v_{p}^{2})+v_{p}^{2}%
}},~~t >t_{p}.  \label{vf}
\end{equation}
The dynamics overall appear much like those of actual particles with non-zero mass.
In contrast, the DWs in a ferromagnet would stop moving as soon as the external torque
ceases.

The AFM DW trajectory in the real space can be subsequently calculated by integrating the
velocity over time. When the time of interest is longer than the pulse duration, the
distance of travel includes the contribution by the inertia given by
\begin{equation}
x_{p}(t )=\frac{1}{4 \lambda}\ln \left( \frac{1+v_{p}}{1-v_{p}}\times \frac{%
1-v_{0}(t )}{1+v_{0}(t )}\right),~~t > t_{p}.  \label{fm}
\end{equation}
Interestingly, this expression also characterizes the length that a DW moving with
the velocity $v_p$  would travel unpropelled before losing all of the energy and
coming to a stop (i.e., via inertia):
\begin{equation}
x_{p}(t \rightarrow \infty )~~ {\Rightarrow }  ~~
x_\mathrm{in}(v_{p})=\frac{1}{4 \lambda}\ln \frac{1+v_{p}}{1-v_{p}}.  \label{vfm}
\end{equation}



Another factor that must be considered in the DW movement is the effect of the
coercivity. In the real structures, the defects like the grain boundaries and impurities may
pin the DW. They keep the DW from slipping via the Brownian thermal motion. Otherwise,
the diffusive dispersion in the DW position would diverge as time $t$ goes to infinity
\cite{Yan2018}. The present formalism can readily account for the non-zero coercivity by introducing
a lower bound to the kinetic energy for displacement. In this picture, the DW gets pinned
as soon as the kinetic energy [$E(v)-E(0)$; see Eq.~(\ref{Ev})] becomes smaller than the
trapping energy $H_{c}M_{L}A \delta_\mathrm{dw}$, where $H_c$ denotes the effective field
for coercivity.  The impact of finite coercivity is particularly
relevant during the inertial phase of the DW transport where the unpropelled movement is
susceptible to the external factors.  While the DW is being driven ($t \leq t_p$) on the
other hand, we may not need an additional consideration so long as the applied SOT is
sufficiently strong.  A corresponding criterion can be expressed in terms
of the critical velocity $v_{c}=\sqrt{4\pi H_{c}M_{L}/\left\vert K\right\vert}$ provided
$v_{c}\ll 1$.
Since $H_{c}$ cannot be determined a priori in terms of the material
parameters, $v_{c}$ is treated as an independent phenomenological constant
for the coercivity of a particular structure.


The expressions given above provide a complete solution to the problem of the DW
dynamics driven by an electric current pulse. Under the pulse
duration $t_{p}$ with amplitude $Q$, the net displacement becomes
\begin{equation}
x_0=x_{Q}(t_{p})+x_\mathrm{in}(v_{p})-x_\mathrm{in}(v_{c}),  \label{fd}
\end{equation}%
where $x_Q (t_p)$ [$=\int_{0}^{t_{p}}v_{Q}(t)dt$] denotes the distance traveled while being driven and
$x_\mathrm{in}(v_{c})$  accounts for the reduction in the free flight (i.e., the inertial
transport) due to the coercivity.  Conversely, they can also be solved to deduce
the relationship between $t_p$ and $Q$ for the desired $x_0$.  It is evident that
a shorter pulse duration is sufficient when the SOT amplitude is stronger and vice versa.


\section{Results and Discussion}

\subsection{Model validation}

Before proceeding further, it is necessary to examined the validity of the adopted
treatment based on the soliton description [Eq.~(\ref{be})]. For this, a direct comparison
is made with the micromagnetic simulation of the AFM DW dynamics \cite{Li2017}.
The calculations are carried out for the AFM slab with the dimensions of 200$\times $50$%
\times $20 nm$^{3}$. The numerical values assumed for the parameters are the sublattice
magnetization $ M_{L}=4\pi \times 320$ G,
the anisotropy energy $K= -1.6\times 10^{6}$ erg$\cdot $cm$^{-3}$, the resonant
frequency $\omega _{r}= 2 \pi \times 122$ GHz, and the damping parameter $\lambda =0.086$.
In addition, the exchange stiffness $A_\mathrm{ex}$ of $10^{-7}$~erg/cm is used that further specifies the
magnon velocity $v_{m}$ of $6.9\times 10^{5}$ cm/s ($= \gamma \sqrt{A_\mathrm{ex} H_\mathrm{ex}/ M _L}$)
\cite{Gomonay2016}. The relation between the SOT $Q$ and the current density $J$ (i.e., $Q/J$)
is estimated to be $0.114$ nm$^{2}$/nA with the effective spin-Hall angle of 0.1 \cite{Shiino2016}.
The effect of coercivity is not considered for the comparison due to the limitation of the
micromagnetic simulation.

The calculations reveal the non-linear dependence of the DW velocity on the pulse
intensity.  As shown in Fig.~2(a), the AFM DW when subjected to an SOT pulse undergoes
acceleration initially and then velocity saturation with the dissipation compensating
the anti-damping SOT.  Once the driving force is turn off, the inertial phase follows.
Using the dimensional units, both the analytical solution and the results of the micromagnetic
simulations are plotted in Fig.~2(b,c). A good agreement is observed between the two approaches
in both the steady-state transport and the transient conditions with the inertial motion
[Figs.~2(b) and 2(b), respectively],
providing credence to the validity of the developed model. With the soliton ansatz verified, Eq.~(\ref{ve})
can be expanded to account for the stochastic thermal motions in the DW dynamics at finite temperatures.

\subsection{Effects of random thermal motion}

The analysis of the thermal field influence on the DW transport supposes evaluation of
thermal fluctuations $h_\mathrm{th}$ along the entire DW path. On the other hand, the
perturbations away from the location of the wall texture are unlikely to affect the DW
dynamics.  The actual range of the AFM channel where the influence of random motions
needs to be considered is the relatively narrow stretch corresponding to
the wall texture (i.e., the DW width $\delta_\mathrm{dw}$).
Thus the problem can be approximated to the analysis of the fluctuation
effect in the finite volume of $V=A\times \delta_\mathrm{dw}$ associated with
the soliton representation. Accordingly, the influence of the thermal field can be
accounted for by conveniently adding of a randomly fluctuating field-like
torque [i.e., $-\frac{\pi}{2 \lambda} \frac{d}{dt }h_\mathrm{th}(t )$] to
the current induced SOT $Q$ in the dynamical equation governing the soliton motion
[see also Eq.~(\ref{rhp})].

In describing the thermal field $h_{th} (t)$, the approximation based on a series
of random step functions used commonly in the ferromagnetic systems cannot be
applied here due to the explicit dependence on the time derivative \cite{Tsiantos2003}.
As an alternative, a spectral representation is adopted in the form of a Fourier
series expansion with random amplitudes \cite{Semenov2018}.  This representation
allows straightforward introduction of the upper and lower bounds in the noise spectrum
by considering the auto-correlation time $\tau_c$ and the characteristic N\'{e}el-vector
relaxation time $\tau_m$ (more precisely, the inverses $2 \pi /\tau_c$ and
$2 \pi / \tau_m$, respectively). Furthermore, the fact that the lower truncation
frequency $2\pi / \tau_m$ corresponds to the broadening of the AFM resonant frequency
$\delta_r$ offers a physical ground for the discretization of the spectral domain in
the comparable intervals. As the response of a damped N\'{e}el-vector motion becomes
practically invariant to the perturbation frequency swing in the range of $\delta_r$
due to the broadening, the actual noise spectrum can be discretized likewise.

In other words, the AFM response to the thermal noise is virtually
equivalent to a series of sinusoidal perturbations with random amplitudes and
the frequencies $n\delta_r$ ($n=1,2,...,N$, where $N$ is given by $ \tau_m/\tau_c$).
The fluctuation dissipation theorem defines the amplitude of fluctuating field in the form
\cite{Semenov2018},
\begin{equation}
h_\mathrm{th}(t)=\frac{\delta _{r}}{\omega _{r}}\left( \frac{2k_BT}{NV\left\vert K\right\vert }%
\right) ^{1/2}\left( \sum_{n=1}^{N}\alpha _{n}\sin n\frac{\delta _{r}}{%
\omega _{r}} t +\sum_{n=1}^{N}\beta _{n}\cos n\frac{\delta
_{r}}{\omega _{r}}t \right) ,  \label{ht}
\end{equation}%
where $k_BT$ denotes the thermal energy, $n\frac{\delta_r}{\omega_r} t $ is actually
$n\delta_r t$ in physical units, and $\frac{1}{2N}\sum_{n=1}^{N}\left\langle \alpha _{n}^{2}
+\beta_{n}^{2}\right\rangle =1$.  Note that the noise expression applies only for
a duration up to $\tau_m$ in the time domain due to the relaxation.
A time period longer than this interval requires repeated random selections.
Equation~(\ref{ht}) is clearly differentiable that can be directly
incorporated into Eq.~(\ref{ve}). The exact details of the noise model including a
particular choice of the material parameters are not highly crucial in examining
the possible electrical control in the thermally induced dispersion of the DW position.

Before accounting for the thermal component, Fig.~3 first examines the deterministic relation
between the necessary SOT pulse strength and the duration for a desired travel distance $x_0$.
The results are plotted in a normalized form in terms of $2 \lambda t_p$ and $2 \lambda x_0$
since it conveniently circumvents the explicit dependence on other physical parameters
[see, for instance, Eqs.~(\ref{vt}) and (\ref{vfm})]. The coercivity of $v_{c}=0.04$ is considered
as well to reflect the conditions in the realistic structures [Eq.~(\ref{fd})].  As shown, both short and
strong (red) as well as long and weak (blue) pulses can shift the DW by the same distance.
The obtained relation serves as the guideline for the desired mean or average motion of the AFM DWs.

Once the influence of thermal fluctuations is accounted for, the DW dynamics deviates from the
prescribed path [see the inset to Fig.~4(a)]. The resulting dispersion in the final position
$x_\mathrm{dw}$ around $x_0$ can be calculated by numerically solving Eq.~(\ref{ve}) in the
presence of the random field-like torque $-\frac{\pi}{2 \lambda} \frac{d}{dt }h_\mathrm{th}(t)$.
A sufficiently large number of iterations $N_i$ are needed to ensure a statistically reliable
outcome due to the stochastic nature of the calculation.
For the numerical evaluation, the parameter
values adopted earlier for Fig.~3 are also applied except $\lambda$, for which
a slightly larger choice of 0.135 (thus, $\delta _{r}=2\pi \times $16.5 GHz) is used here.
Likewise the magnitude of the auto-correlation time $\tau _{c}$
is treated empirically.  Since our analysis is not significantly affected by the exact value
of $\tau_c$ so long as it is sufficiently shorter than $\tau_m$, a small constant fraction
($ \tau_c = 0.01 \tau_m$; $N=100$) is assumed for simplicity in Eq.~(\ref{ht}). For a set of
given conditions, the calculations are repeated 250 times ($=N_i $), each with an independently
selected $h_\mathrm{th}(t)$ pattern randomly varying in time.

Figure~4(a) shows the typical probability distribution of the DW position for two SOT pulses at 300 K.
While both pulses are designed to shift the DW by the same distance on average (see also
two points marked in Fig.~3), the dispersion as the result of random thermal motion shows
a significant difference.  The short and strong pulse (red; $Q=0.015$) produces a much narrower
distribution than the long and weak counterpart (blue; $Q=0.007$).  It can be intuitively understood
that the DW driven by a strong pulse is less likely to be influenced by the comparatively minor thermal
fluctuations. At the same time, the DW  is exposed to the random motions for a shorter duration in this case for
it reaches the final destination quicker and then pinned by the coercivity. It is also interesting
to note that the mean position $\left\langle x_\mathrm{dw} \right\rangle $ averaged
over $N_i$ iterations is indeed given
by $x_0$ as designed (from the deterministic analysis), with high accuracy. Since the pulse strength
and duration can be controlled electrically, the probability distribution function in the DW position
can be tuned likewise (i.e., narrower $\rightleftarrows$ broader).  Combined with the structure utilizing
the anisotropic magnetoresistance as shown in Fig.~1(a) \cite{Yu2018,Park2011,Lu2018}, it can be translated
into a corresponding probabilistic
distribution in the electrical output signal $-$ an essential component in the probabilistic computing or
Bayesian learning.  Finally, Fig.~4(b) plots the normalized standard deviation in the DW position
$\Delta =\left\langle (x_\mathrm{dw} /\left\langle x_\mathrm{dw} \right\rangle -1)^{2}\right\rangle ^{1/2}$
as a function of $Q$ for three average displacements. The observed large differences in $\Delta$
indicate that a broad range of probability distributions can be realized with a single AFM device.
Moreover, the normalized deviation appears to be only weakly dependent on $x_{0}$
or $\left\langle x_\mathrm{dw} \right\rangle $ (thus, on the pulse duration $t_p$ for a given $Q$). Clearly,
the determining factor is the pulse strength $Q$, to which $\Delta$ exhibits an inverse proportionality.

\section{Summary}
The fluctuating thermal fields in the antiferromagnets are identified as a potential mechanism to realize
a probabilistic distribution in the electrical output signal, whose characteristic properties such as the standard deviation
may also be tailored conveniently by electrical control. To this end, the dynamics of a 90$^{\circ }$ DW driven by the current-induced SOT is theoretically examined at finite temperatures in a thin-film AFM structure. A soliton-type representation based on an energy balance equation is developed to describe the DW motion in combination with a stochastic thermal field model for AFM nano-particles.  The calculation results clearly illustrate that both the average displacement of the DW position and its thermally induced dispersion can be modulated electrically by simply tuning the driving SOT strength and the duration.  The corresponding probabilistic response in the electrical output signal via the magnetoresistance offers an effective means to realize the probability distribution functions that can be "trained".  This unusual functionality may provide a key component in the emerging probabilistic computing and machine learning architectures.

\begin{acknowledgments}
This work was supported, in part, by the US Army Research Office (W911NF-16-1-0472).
\end{acknowledgments}

\clearpage

\begin{center}
\begin{figure}[tbp]
\includegraphics[width=8cm,angle=0]{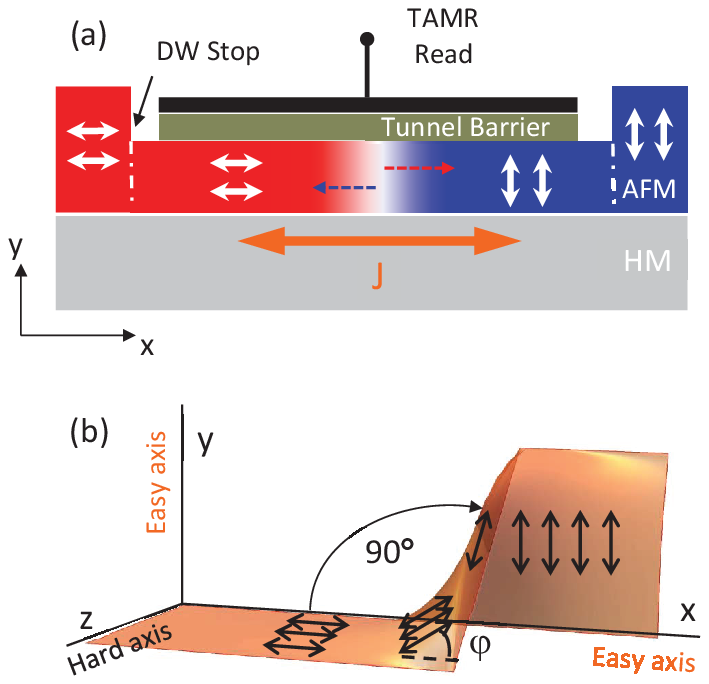}
\caption{(a) Schematic illustration of the antiferromagnet-based DW device under consideration.
The electric current $J$ through the heavy metal (HM) layer generates the spin current, providing the
SOT to the AFM strip necessary for the DW motion.  The 90$^\circ$ DW can travel in either direction ($\pm x$) that
is determined by the current flow in the HM layer.  The shift in the DW position
can be detected through the top contact either via the anisotropic magnetoresistance or its tunneling variety (TAMR) as shown.
(b) Spin textures of the $90^{\circ}$ N\'{e}el DW with the heavy $z$-axis and the easy $x$- and $y$-axes.}
\end{figure}
\end{center}

\clearpage

\begin{center}
\begin{figure}[tbp]
\includegraphics[width=14cm,angle=0]{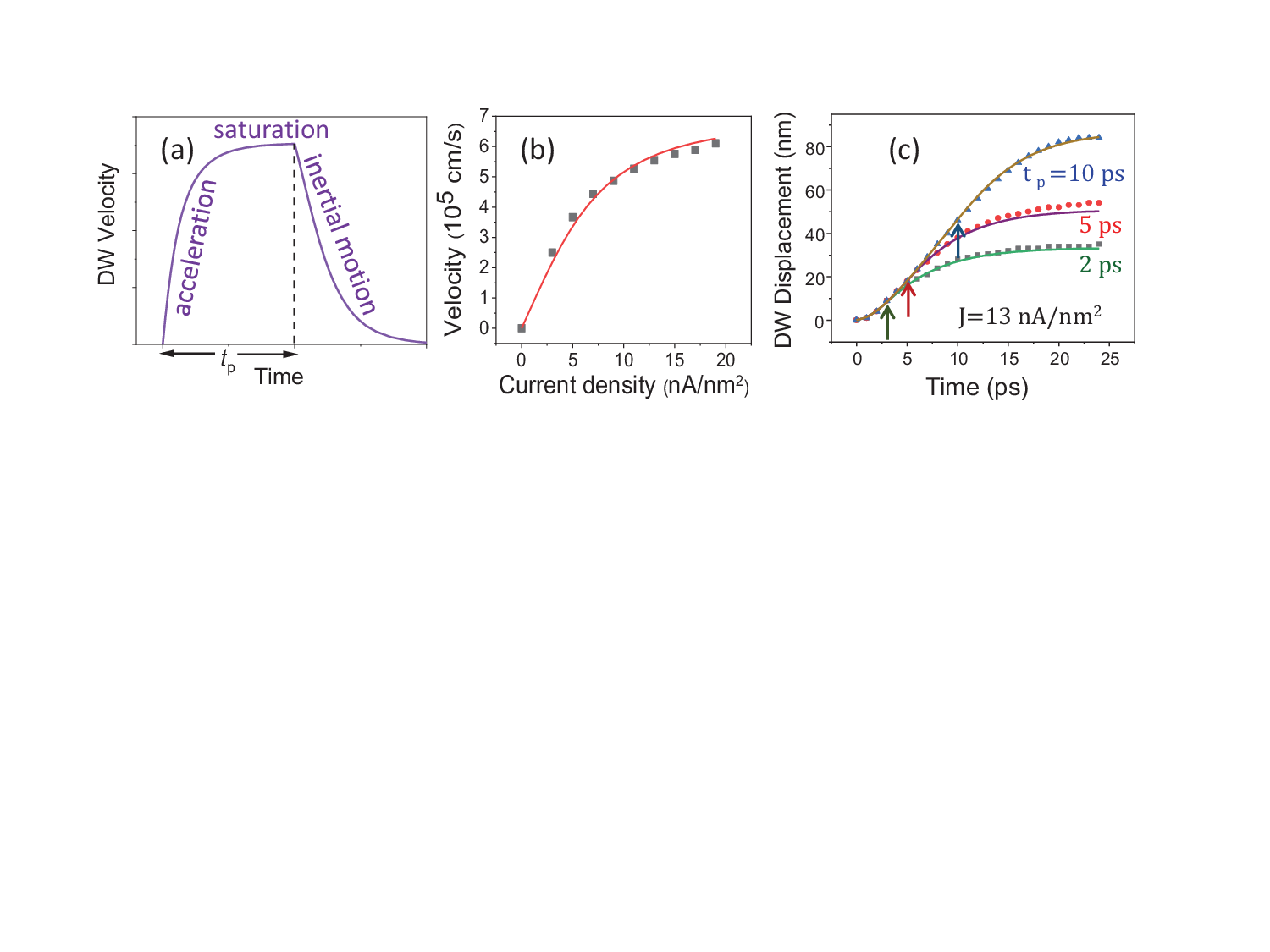}
\caption{Comparison of the calculation results with the micromagnetic simulations in the deterministic transport
(i.e., without thermal motions).  (a) Characteristic evolution of the 90$^\circ$ DW velocity in the thin-film AFM strip
under a SOT pulse with duration $t_p$.
(b) Steady-state DW velocity vs. the driving current density in the HM layer.  (c) DW position as a function of
time for different SOT pulse durations. The vertical arrows indicates the instants when the SOT is turned off.
The driving current density $J$ is fixed at 13 nA/m$^2$.
In (b,c), the data points are from micromagnetic simulations while the solid curves show the results
from the model calculations.  An excellent agreement is observed between them in both the steady-state and transient
transport.  The effect of coercivity is not considered here.
}
\end{figure}
\end{center}

\clearpage

\begin{center}
\begin{figure}[tbp]
\includegraphics[width=8cm,angle=0]{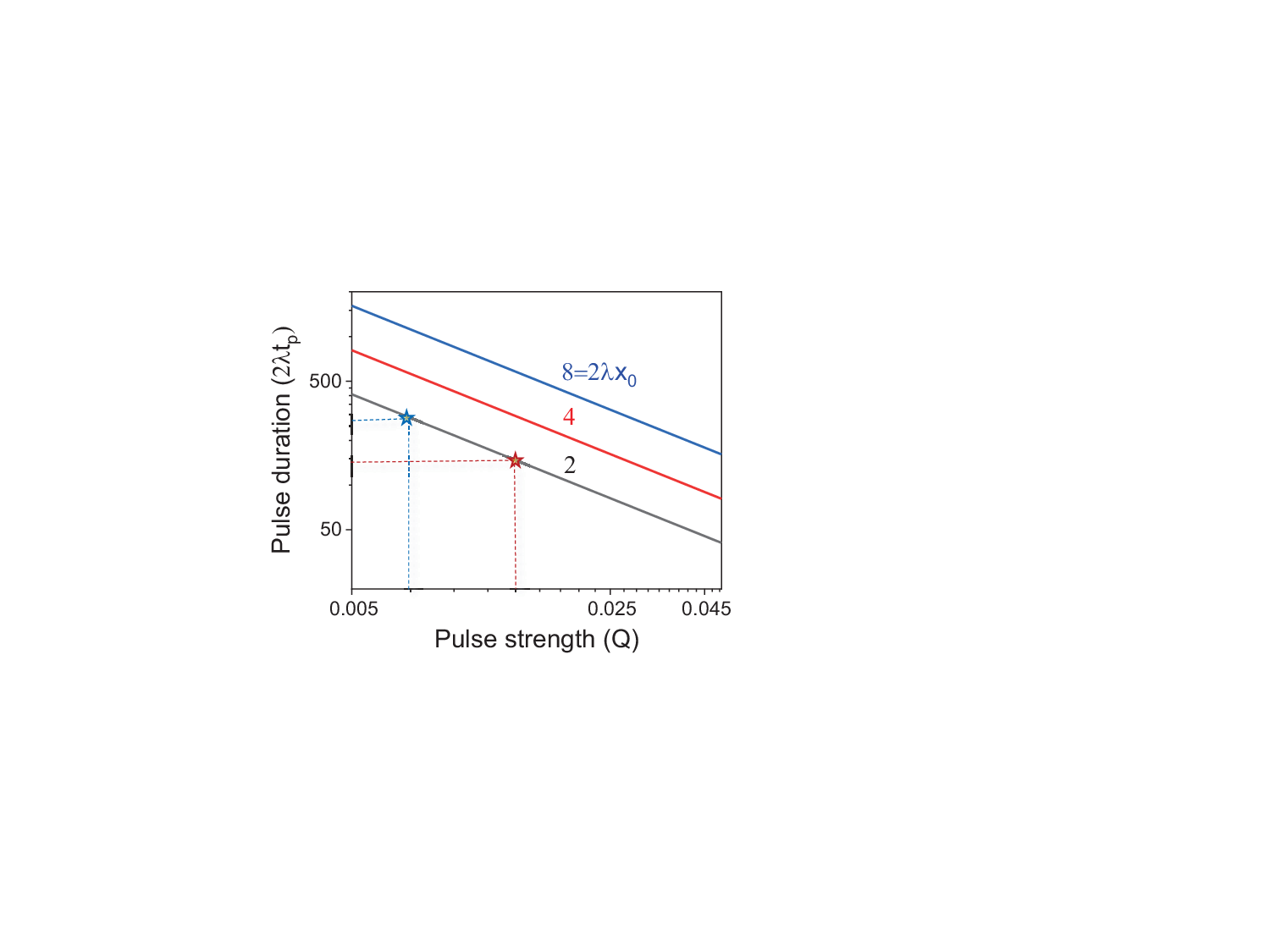}
\caption{Calculated SOT pulse strength vs.\ the duration for three travel distances
$x_0$ in the deterministic transport. The results are plotted in normalized units in terms of
$2 \lambda t_p$ and $ 2 \lambda x_0$ to circumvent the explicit dependence on other physical parameters.
The coercivity of $v_c=0.04$ is also included in the analysis. The blue and red points highlight two different
conditions ($Q =$ 0.007 and 0.015, respectively) that shift the DW by the same distance.}
\end{figure}
\end{center}

\clearpage

\begin{center}
\begin{figure}[tbp]
\includegraphics[width=12cm,angle=0]{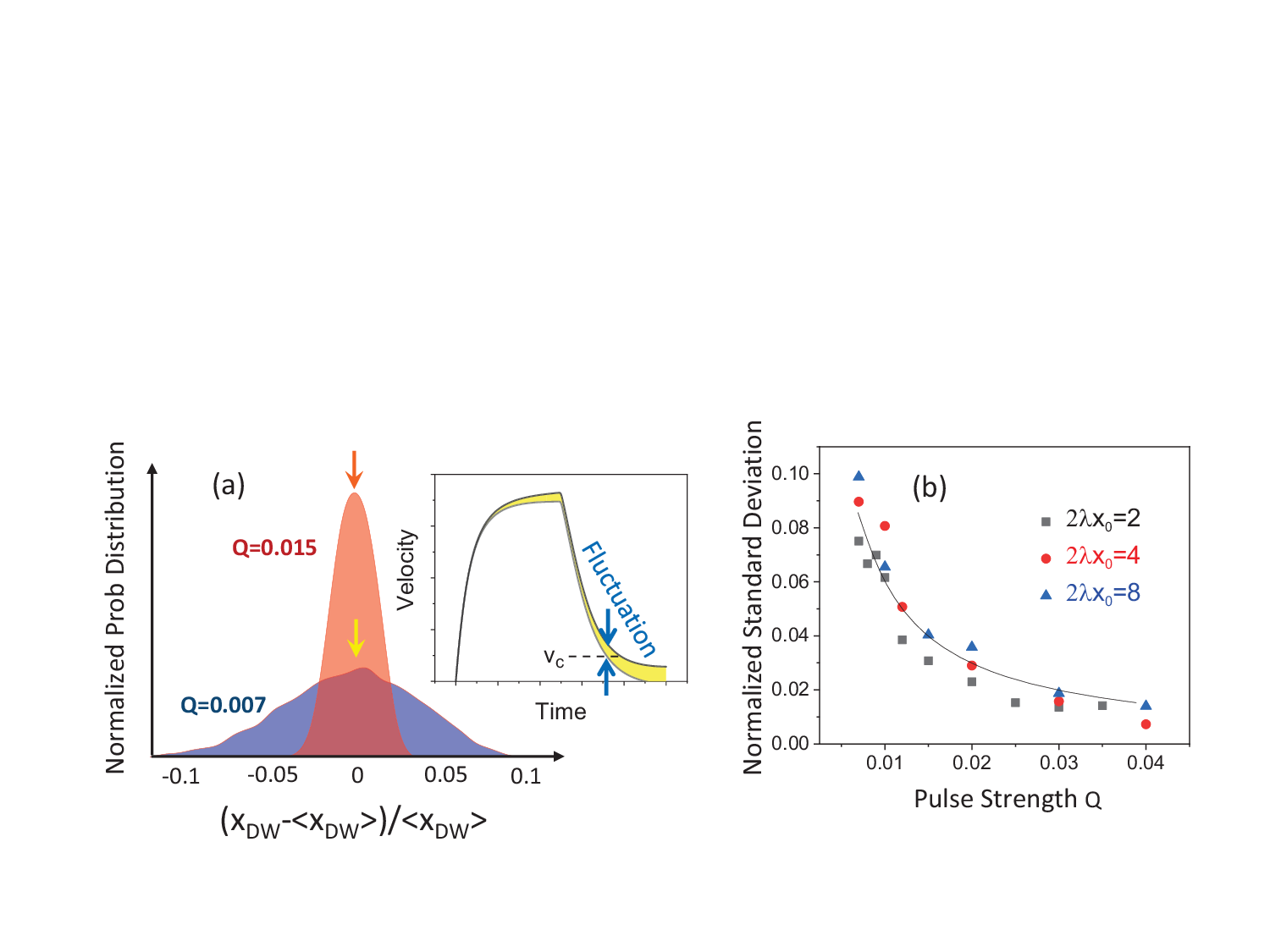}
\caption{(a) Probability distribution of the DW position $ x_\mathrm{dw}$ at 300 K
calculated for two SOT pulses ($Q=0.015$ and $ Q=0.007$ corresponding to the red and blue
marks in Fig.~3, respectively). The horizontal axis shows the normalized deviation from the average
position $\left\langle x_\mathrm{dw} \right\rangle$ which matches with $x_0$ from the deterministic analysis.
The inset schematically illustrates the effect of random thermal fluctuations on the DW velocity.
(b) Normalized standard deviation in the DW position as a function of the SOT strength for three
average displacements $x_0$. The pulse duration corresponding to each $Q$ and $x_0$ can be found in Fig.~3.
The solid curve represents a fit with an inverse proportionality $aQ^{-1}$
($a=6\times10^{-4}$). The coercivity ($v_c=0.04$) is accounted for in both (a) and (b).
}
\end{figure}
\end{center}

\end{document}